Title: True Polar Wander of Enceladus From Topographic Data.

Authors: Radwan Tajeddine[1], Krista M. Soderlund[2], Peter C. Thomas[1], Paul Helfenstein[1], Matthew M. Hedman[3], Joseph A. Burns[4], Paul M. Schenk[5]

[1]Center for Astrophysics and Planetary Science, Cornell University, Ithaca, NY 14853, USA
[2]Institute for Geophysics, John A. & Katherine G. Jackson School of Geosciences, The University of Texas at Austin, Austin, Texas 78758-4445, USA
[3]Department of Physics, University of Idaho, Moscow, ID 83844-0903, USA
[4]College of Engineering, Cornell University, Ithaca, NY 14853 USA
[5]Lunar and Planetary Institute, Houston, TX 77058, USA

**Abstract**

Many objects in the solar system are suspected to have experienced reorientation of their spin axes. As their rotation rates are slow and their shapes are nearly spherical, the formation of mass anomalies, by either endogenic or exogenic processes, can change objects' moments of inertia. Therefore, the objects reorient to align their largest moment of inertia with their spin axis. Such a phenomenon is called *True Polar Wander* (TPW).

Here we report the discovery of a global series of topographic lows on Saturn's satellite Enceladus that we interpret to show that this synchronously locked moon has undergone TPW by ~55° about the tidal axis. We use improved topographic data from the spherical harmonic expansion of Cassini limb and stereogrammetric measurements to characterize regional topography over the surface of Enceladus. We identify a group of nearly antipodal basins orthogonal to a topographic basin chain tracing a non-equatorial circumglobal belt across Enceladus' surface. We argue that the belt and the antipodal regions are fossil remnants of an earlier equator and poles, respectively. We argue that these lows arise from isostasic compensation and that their pattern reflects spatial variations in internal dynamics of the ice shell. Our hypothesis is consistent with a variety of geological features visible in Cassini images.

1. Introduction

Besides the relative motion of lithospheric plates, the Earth as a whole moves with respect to its rotation pole, as shown by paleomagnetic, astrometric, and geodetic measurements (Mitrovica & Wahr, 2011). Such True Polar Wander (*TPW*) occurs because our planet's moments of inertia change owing partly to internal thermal convection (Doubrovine et al. 2012). Thus, to conserve angular momentum while losing rotational energy, Earth's axis of maximum moment of inertia aligns with its spin axis. Similar reorientations have been proposed for other celestial bodies (e.g. Schenk et al. 2008; Bouley et al. 2016; Siegler et al. 2016), although supporting evidence is often fragmentary (Matsuyama et al. 2014). Evidence of such a phenomenon significantly improves interpretations of the body's geophysical evolution.

Among those bodies is Saturn's moon Enceladus; almost as remarkable as the geyser activity on the satellite is the location of these hotspots at the South Pole in a ~400-m deep topographic depression (Porco et al. 2006; Spencer et al. 2006). To explain this placement, some studies (Nimmo & Pappalardo, 2006; Matsuyama & Nimmo, 2008) suggest that the hotspot moved to the polar region after forming elsewhere. Indeed, various aspects of Enceladus' surface geology could reflect changes in the satellite's orientation (Crow-Willard & Pappalardo, 2015; Spencer et al. 2009; Helfenstein et al. 2010; Nahm & Kattenhorn, 2015).

To further investigate the hypothesis of TPW on Enceladus, we determine its global shape. Past studies of the satellite's shape include Thomas et al. (2007) and Thomas (2010), who used limb profile data from the Cassini Imaging Science Subsystem (ISS) to fit an ellipsoid to the shape of Enceladus. They concluded that its triaxial shape is not consistent with an object in hydrostatic equilibrium. Degree eight coefficients of the spherical harmonics function (Nimmo et al. 2011) with non-relaxed topographic variations up to 1 km, show more evidence of the non-equipotential surface of Enceladus. These topographic variations were also reported by Schenk & McKinnon (2009) who used the method of stereogrammetry, focusing in particular on large basins spread over Enceladus' surface; they suggested isostatic compensation as a potential mechanism to form these basins. However, their map covered only 50% of Enceladus' surface and could not support a conclusion as to whether these basins were randomly distributed or were part of a global pattern. In this work, we fit the coefficients of the spherical harmonics function up to degree 16 by combining updated positions of limb profiles and control point measurements from Cassini images.

In section 2, we describe the different types of Cassini ISS data that were used in this work to fit the shape of Enceladus. In section 3, we introduce the spherical harmonics function, with initial fits and comparisons to previous works. We extend the fit of the spherical harmonics function to a higher order in section 4, report a set of basins on Enceladus' surface, and interpret them in terms of True Polar Wander. We offer a geophysical interpretation of the basins in section 5, consistent with the TPW hypothesis. Geological evidence in section 6 further supports our hypothesis and geophysical interpretations. We give a summary and conclusions in section 7.

**2. Data**

Our global shape model is derived from a combination of limb profiles and control points. The improved coverage of surface features along limb lines complements the more homogeneous coverage on the satellite's surface provided by control points. Our expanded data set has 54 limb profiles that contain surface coordinates of 41780 points as well as 6245 stereogrammetrically derived control points (Thomas et al. 2016; Edwards, 1987; Schenk & McKinnon, 2009).

**2.1 Limb profiles**

Besides spacecraft astrometry (Tajeddine et al. 2013, 2015; Cooper et al. 2014) limb measurements are used to study shapes and topography of planetary bodies. Limb-coordinate measurement techniques are described in Thomas et al. (1998) and Thomas et al. (2007). Limb-finding involves sub-pixel modeling of the bright edge of an illuminated object. The precision of limb-coordinate measurements can reach better than 0.1 pixels. Accuracies, checked by images of greatly differing resolutions or with independent data, are generally ~0.15 pixels. The solutions for shapes are affected by how well an ellipse centre is fit for each view; the rougher the object or the shorter the limb arc, the less accurate the calculated centre and the less accurate the overall ellipsoidal solution. Views with more than 180° of limb arc, such as those obtained during transits of the Saturnian disk, those at very low phase, or those with Saturn shine fix centres with very high accuracy and thereby help stabilize centres of all other views. Uncertainties in centre locations are obtained by removing part of the data from the ends of limbs arcs (10%) and finding the differences in fit centres between the full and truncated data. The geometry of projected limbs of ellipsoids is described in detail in Dermott & Thomas (1988). It may be argued that topographic variations may induce biases in the limb-points' coordinates. For instance, a mountain located in front of --or behind-- the actual location of the limb will induce an error in the position of that mountain. However, tests on synthetic data (Nimmo et al. 2010) indicate that these biases do not affect the global scale long-wavelength topography, although they may affect positions of small-scale topographies. The uncertainty on the positions of limb points ranges between 50 m and 1.67 km depending on image resolution, with a mean value of 270 m.

**2.2 Control points**

Control points have been generally used to study the rotation of planetary satellites (Thomas et al. 2016; Tajeddine et al. 2014; Oberst et al. 2014). They have been used in some cases to build shapes of bodies where limb profiles were unavailable, like the case of Titan (Zebker et al. 2009); however, this study is the first time that limb profiles are combined with control points to build the shape of an object. Here we used them to build a refined shape of Enceladus. Measurements of control points have been done using both manual and automated methods.

In the first method, control points are manually digitized surface features, mostly craters. Techniques used here rely on basics discussed in Davies et al. (1998) and Thomas et al. (2002). Image coordinates of surface control points are rotated with the camera's inertial orientation (C-matrix), scaled by the camera's optical parameters in combination with the relative positions of target and spacecraft, to provide body-centred vectors. The array of these observed image coordinates is then fit to predicted coordinates in the target body's coordinate frame; measurement in at least three images with convergence angles >10° is required. Most of the software used in this work was developed by J. Joseph (Thomas et al. 2002) for the NEAR mission with subsequent modifications by B. Carcich and J. Joseph. This method provided 487 coordinates of surface control points.
In the second method, the USGS-ISIS2 software (Edwards, 1987; https://isis.astrogeology.usgs.gov/) was used for automatic recognition of surface features and photogrammetric reconstruction of their coordinates. Each surface control point requires two observations from two different viewing angles for triangulation. A total number of 5758 surface coordinates were acquired with this method.

The Cassini camera's optical parameters (focal length, distortion) are sufficiently accurate that they introduce errors much less than 0.1 pixels across the detector; calibration of the ISS Narrow and Wide-Angle cameras (NAC, WAC) is described in (Owen, 2002), the geometric portion of which is based on in-flight stellar images (West et al. 2010). The NAC provides scales of 6 μrad/pixel (6 km/pixel at $10^6$ km range), and the WAC provides 60 μrad/pixel (60 km/pixel at $10^6$ km range). Fields of view of the two cameras are 0.35° and 3.5°, respectively.

All images require pointing corrections because achievable precision in the measurements is far better than the camera pointing information. In this operation, the target body's center is shifted in line and sample (X, Y). We do not generally allow the twist (rotation about the optical axis) to vary if the solution has any rotational outcome of interest. Because image pointing is allowed to change, the residuals in the images are determined by the relative spacing of the projections of the points in the image, rather than by total rotational offsets. Thus, for each solution, all of the body-centered positions in each image are recalculated, and a change of any input data or assumed spacecraft position (including the rotation model) can affect all computed body-centered X, Y, Z positions. Average residuals generally approach 0.3-0.45 pixels, and the uncertainty on the reconstructed positions of control points ranges between 30 m to 650 m depending on image resolution and number of observations par point, with a mean value of 155 m.

**3. Shape model.**

The shape of a solar system body can be represented by expressing the radii at various locations by a series of spherical harmonics function, where the radius is given as,

$$r(\theta,\phi) = R_0 \sum_{l=0}^{N} \sum_{m=0}^{l} P_{lm}(\cos\theta)\left[C_{lm}\cos(m\phi) + S_{lm}\sin(m\phi)\right] , \qquad (1)$$

where $R_0$ is the body's mean radius, and $P_{lm}$ is an associated Legendre polynomial of degree $l$ and order $m$. $\theta$ and $\phi$ are the colatitude and longitude, respectively. $C_{lm}$ and $S_{lm}$ are the coefficients that represent the deviation from the sphere of radius $R_0$. For instance, the shape at order $l=0$ is a sphere where $C_{00}$ adjusts the mean radius from $R_0$, $C_{1m} = 0$ if the coordinate origin is at the figure's center, $l=2$ involves equatorial and polar flattening and represents an ellipsoid, $l=3$ represents the asymmetry between the hemispheres, and so on… The higher the function goes in degree of spherical harmonics, the more it represents local topographic variations. The degree to which the shape fit is reached depends on the amount of available data.

Each of the (X,Y,Z) coordinates of the data points has been converted to longitude ($\phi$), colatitude ($\theta$), and radius ($r$). Then, we fit the coefficients $C_{lm}$ and $S_{lm}$ in Eq. 1, applying a Weighted Least Squares method, where we minimize the $\chi^2$ residuals between the observed radii (from data points) and the calculated ones at the same spherical coordinates ($\phi$, $\theta$) (starting with a sphere of radius $R_0 = 252.1$ km; Thomas, 2010; and setting initial values of $C_{lm} = 0$, and $S_{lm} = 0$, except for $C_{00} = 1$) weighted by the estimated uncertainty on the position of each data point (see section 2). Although the number of data points coming from limb profiles is 10 times larger than that coming from control points, the latter offer a more homogeneous coverage of the satellite's surface, which fills in the gaps of uncovered surface with the limb profile data (see Fig. C1).

We first tested our method by performing fits to a low degree in spherical harmonics arriving at a mean radius $R_0$ = 252.22 km, and ellipsoidal axes $a$ = 256.53 km, $b$ = 251.45 km, and $c$ = 248.66 km, which match the currently accepted ellipsoidal shape of Enceladus (Thomas et al. 2016). Next, we fit the shape to order eight in spherical harmonics for comparison with previous work (Nimmo et al. 2011). We obtain a similar global map for $l$=3 to $l$=8 topography (Fig. C1) with minor differences due to our improved coverage; particularly, the regions at (90°W, -45°N) and (30°W, 45°N) appear to be less mountainous than previously published in Nimmo et al. (2011). We also confirm the satellite's polar asymmetry reflected mainly by the coefficient $C_{30}$ (Thomas et al. 2007, Nimmo et al. 2011).

## 4. Higher order shape model and evidence of TPW

The amount of available data allowed us to model the shape of Enceladus to degree 16 in spherical harmonics. By removing the $l$=0-2 terms, one sees the topographic variations on top of the average ellipsoid. Figure C2 in the extended data shows the topographic map of Enceladus represented as $l$=3 to $l$=16 in spherical harmonics. However, the topography of Enceladus is dominated by the coefficients representing the polar asymmetry ($C_{30}$, $C_{50}$, $C_{70}$, etc.); also, the SPT is the youngest feature on Enceladus' surface, and it is very likely that other topographic features have formed earlier. Therefore, to appreciate the older topography superposed on this dominant feature, we plot a topographic map subtracting those terms from the spherical harmonic function (Fig.1a). It reveals a very different class of structures: eight 90-130 km wide basins ($E1$-$E8$) with an average depth of 0.7 km that form a circumglobal basin chain approximately following a non-equatorial great circle that is approximately aligned with Enceladus' cratered terrains (Fig.1b). Moreover, two basin groups of 0.5-km average depth are scattered around nearly antipodal locations in Enceladus' resurfaced trailing and leading hemisphere terrains (Fig.1b): basins $S1$-$S3$ lie around average coordinates (272°W, -27°N) and basins $N1$ and $N2$ lie around average coordinates (79°W, 25°N). The basin centered at (180°W, -75°N) is a remnant of the South Polar Terrain (SPT) that was not removed, when subtracting the polar asymmetry features, due to topographic asymmetry relative to the spin axis; we note another deep basin centered at (230°W, -15°N) that is not part of any group of basins, which may have a different geophysical origin than the E, S, and N basins. In comparison to the incomplete stereographic map of Schenk & McKinnon (2009), we note the similarities between our map and theirs. For example, our basins E1, E4, E5, E6 are the same ones denoted E, C (extended to B), A, and F (extended to D), respectively, in their map. Their map also shows parts of the S1-3 basins. Giese et al. (2010) found similar basins in their incomplete topographic map.

All of the basins that we report here are not a consequence of our higher order terms ($l$=16); rather, they start appearing at $l$=7 and become increasingly better defined at higher orders. While the shapes of these basins vary, the global picture remains the same, implying that these basins are not an artefact of the spherical harmonics function.

Furthermore, error analysis suggests a 3σ uncertainty on topography below 100 m (Appendix A). Thus, all of the reported basins are statistically significant.

We fit a plane to the circumglobal set of $E$ basins to characterize the geometry of the lows (red line on Fig. 1a); the normal to the plane crosses the surface at points $\bar{N}$ (79°W, 35°N) and $\bar{S}$ (259°W, -35°N), putting $\bar{N}$ near the centre of $N$ points and $\bar{S}$ at the southern end of $S$ points. The average arc distance from $\bar{N}$ to $E$ points is 82°, to $N$ points 10°, and to $S$ points 166° (Fig.1b, Table 1), noting that these are distances to centers of basins and that each basin has a width of ~20-30°. In addition, the $S$ and $N$ basins are spread over ~45-60° on Enceladus' surface (Fig.1a), similar to the extent of today's SPT.

Note that our detection of the alignment of the $E$ basins was done by eye. To evaluate the quality of this detection we tested whether different configurations of basins could align on different great circles (see Appendix D). Our testing criteria are: (1) the average depth along the great circle, and (2) whether other groups of basins gather around the poles. For that we tested different pole positions with longitude between 0° and 360°, and colatitudes between 0° and 90° (limited to this range to avoid repetition due to hemispheric symmetry). While there are two other configuration of basins that could align on a great circle with an average depth within 2σ to 3σ of topography distribution (thin lines in Fig. 1a, and Fig. D2), the proposed $E$ basin alignment is the only one that has an average depth along the great circle beyond 3σ of topography (highs and lows) distribution (Fig. D1). Furthermore, the proposed $E$ basin configuration is the only one that has two groups of basins at its poles (Fig. D2). Therefore, we conclude that the arrangement of $E$, $S$, and $N$ is unique. See Appendix D for more details on the test.

The axes of the fitted plane are not aligned with the directions associated with tidal and rotational distortions. Moreover, many natural satellites are particularly susceptible to polar wander (Matsuyama et al. 2014) because their roughly spherical shapes (i.e. lack of an equatorial bulge due to relatively slow rotation) permit small internal changes to profoundly affect the orientation of the principal axes of inertia. Thus, TPW is an attractive mechanism to interpret the locations of the higher-order surface features described above. Hence, we hypothesize that the satellite's icy shell experienced a reorientation of ~55°±7.5° (3σ, Appendix A2) about an axis that is ~11°±10° (3σ, Appendix A2) east of the sub-Saturnian axis, namely the tidal axis; consistent with the theory suggesting that reorientation is more easily accomplished if it occurs about the tidal axis (Matsuyama et al. 2014).

After backward reorientation of the satellite, the resulting spatial arrangement of terrains (Fig. 2) defines a prominent pattern of polar symmetry that may reveal clues to the geophysical mechanisms that have shaped Enceladus' surface evolution. First, the $E$-basins and most of the old cratered terrains align along an almost circumglobal equatorial band, disrupted by the present SPT that becomes located on the leading hemisphere (Fig. 2a). Second, today's provinces of tectonically resurfaced terrains on the leading (longitude 0-180°W, Fig. 1b) and trailing (longitude 180-360°W, Fig. 1b) hemispheres move after backward reorientation to the North and South Polar regions, respectively, along with N and S basins (Fig. 2b). The average topography as a function of latitude in the reoriented configuration is shown in Fig. D2c. As expected, the lowest mean topographies gather around three latitudes: the poles and the equator. The wide mean topographic low between latitudes -60°N and -30°N (Fig. D2c) contains the current SPT that was not entirely removed when subtracting the terms representing the polar asymmetry as mentioned above, as well as the only basin (out of 13) that does not fit in the big picture of polar symmetry. Interestingly, the lowest mean topography for the *E*-basins, is not perfectly aligned with 0° latitude, but lies approximately 7° to the north. We also note that, on average, the topographic lows are surrounded with topographic highs, particularly, those parallel to the *E*-basins (Fig. 1a). These results raise the question about the origin and arrangement of paleo-equatorial and paleo-polar basins (as well as the smaller mountains).

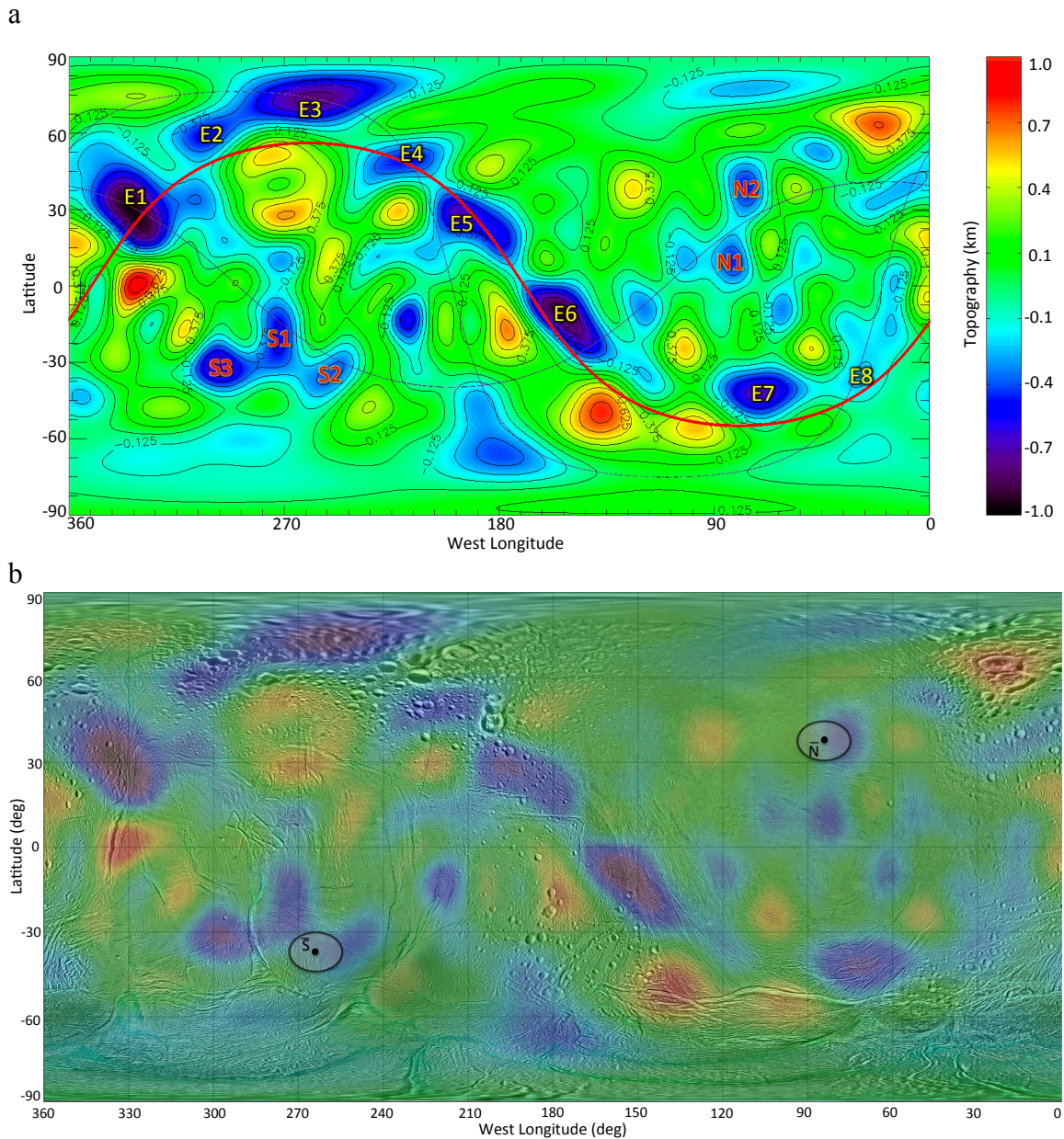

**Figure 1.** Topographic and visible maps of Enceladus. (a) The topographic map is based on the *l*=4 to 16 terms in a spherical harmonic expansion of the satellite's shape. The map reveals series of basins (E1-E8) aligned to form a circumglobal basin chain, and two clusters of basins (S1-S3 and N1, N2) located at nearly antipodal coordinates. The thin blue lines are alignments on different topographic lows that have insignificant average depths (see Appendix D). Contour intervals are 125 m. (b) Overlap of the topographic map over a visible map of Enceladus. $\bar{N}$ and $\bar{S}$ are the poles of the plane fitted to *E* basins. The gray ellipses represent location uncertainties. Graduation of 30°.

| Basin | West longitude (deg) | North latitude (deg) | Depth (km) | Approximate Surface area (km$^2$) | Arc distance from $\bar{N}$ (79°W,35°N) |
|---|---|---|---|---|---|
| E1 | 332 | 28 | 0.9 | 4700 | 85 |
| E2 | 302 | 58 | 0.4 | 960 | 77 |
| E3 | 257 | 72 | 0.7 | 2070 | 69 |
| E4 | 220 | 48 | 0.6 | 2280 | 86 |
| E5 | 196 | 28 | 0.6 | 3430 | 91 |
| E6 | 153 | -12 | 1.0 | 4100 | 86 |
| E7 | 71 | -43 | 0.6 | 2070 | 82 |
| E8 | 15 | -17 | 0.4 | 2740 | 82 |
| S1 | 272 | -18 | 0.6 | 2420 | 156 |
| S2 | 248 | -32 | 0.4 | 900 | 169 |
| S3 | 297 | -31 | 0.6 | 1890 | 148 |
| N1 | 81 | 10 | 0.4 | 1200 | 29 |
| N2 | 77 | 40 | 0.6 | 1970 | 2 |

**Table 1.** Coordinates and depths of the lowest points in the identified basins. $\bar{N}$ represents the north pole of the fitted plane to the *E* basins. The average arc distances from $\bar{N}$ to *N*, *S*, and *E* basins are 22°, 151°, and 82°, respectively. The depths are obtained from a spherical harmonic fit of limb data and control points that do not cover each basin entirely; therefore, some of these basins may be deeper in reality (Schenk & McKinnon, 2009). The surface areas of basins are approximate estimations as their shapes vary depending on the degree of spherical harmonics.

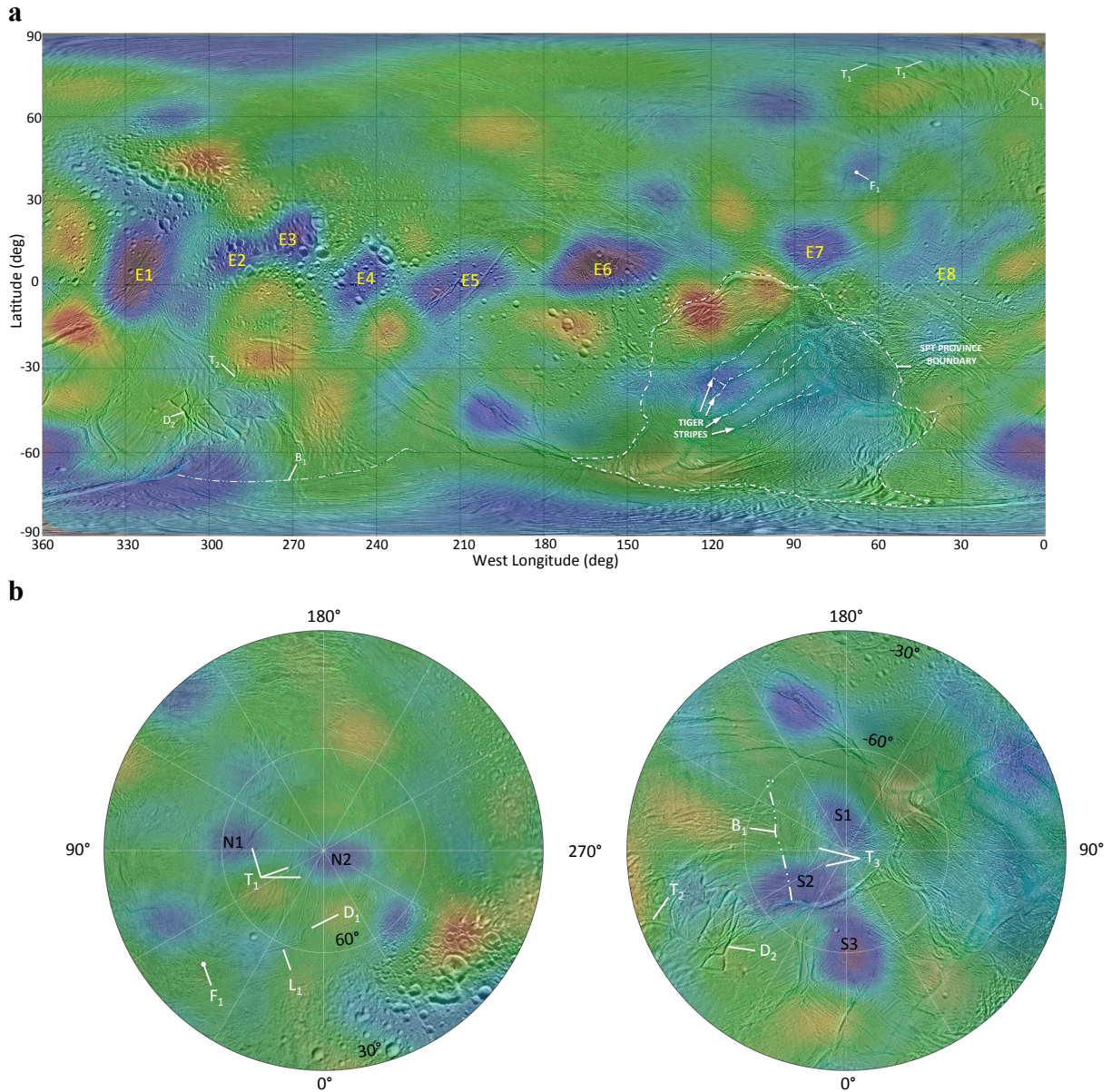

**Figure 2.** Overlap of topographic and visible maps of Enceladus after backward reorientation. (a) Enceladus, as viewed in a cylindrical projection, after its backward reorientation by 55°. The reorientation axis is 11° to the East of today's sub-Saturnian axis. Most of the old cratered terrains lie in a band near the equatorial region along with the circumglobal basin chain, while (b) the young resurfaced terrains (including the *S* and *N* basins) move to the Polar Regions, as shown in the stereographic polar view of the reoriented map. North and South poles are represented by the left and right panels, respectively. The letters B, D, F, and T represent terrains that have similarities with Enceladus' SPT (See text). Graduation of 30°.

## 5. Possible origin of Enceladus' topography

Given the prominence of Enceladus' basins, we focus here on possible mechanisms for their formation and comment briefly on the topographic highs. Topographic depressions have several potential origins: impacts, near-surface geology, geoid variation due to mass anomalies in the deep interior, dynamic topography due to convection, or isostatic compensation due to ice shell thickness or density variations (e.g., Schenk & McKinnon 2009). The basins we identify are not correlated with any known impact features or geologic boundaries, suggesting a deep rather than near-surface origin. Since geoid anomalies are incompatible with the horizontal scale of the basins and dynamic topography associated with convective downwellings is limited to basin depths of approximately 100 m, the most plausible explanation for these basins is isostatic compensation (Schenk & McKinnon 2009; Besserer et al. 2013). Local thinning of the ice shell at the ocean-ice interface or the presence of water lenses within the ice shell introduce positive mass anomalies in a column and depress the surface through Airy isostasy (Fig. 3a). Conversely, mass anomalies owing to local density variations in the shell as a consequence of colder ice or reduced surface porosity may also lower the surface by means of Pratt isostasy (Fig. 3b).

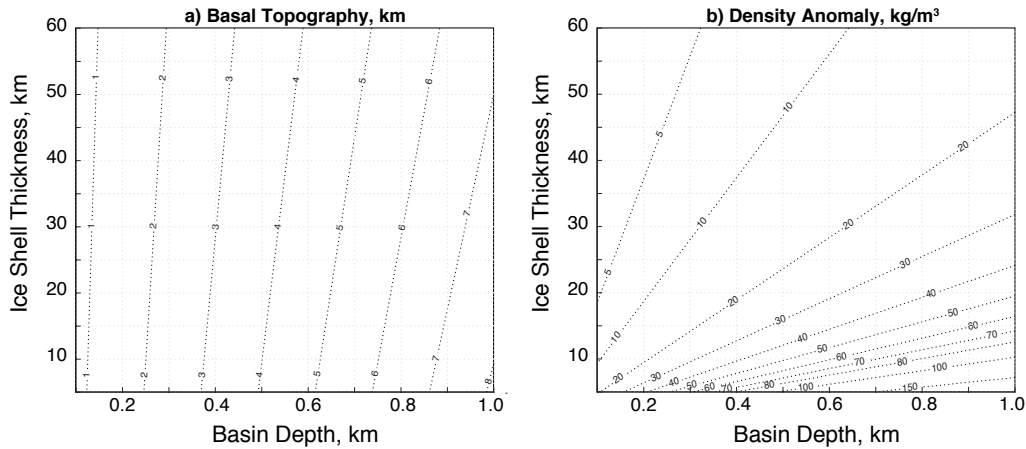

**Figure 3.** (a) Basal topography required to explain surface depressions assuming Airy isostasy with surface ice density of 935 kg/m$^3$ for cold ice, basal ice density of 917 kg/m$^3$ for warm ice, and ocean density of 1030 kg/m$^3$ (Glein et al. 2015; Čadek et al. 2016). (b) Local density anomaly required to explain surface depressions assuming Pratt isostasy with a mean ice shell density of 925 kg/m$^3$. Calculations follow the approach of Schenk and McKinnon (2009).

As shown in Fig. 3, the *E* basins, with average depths of 0.7 km, require basal upwarping of approximately 5 km or a density anomaly of 11 to 150 kg/m$^3$ with the lower (upper) bound corresponding to a 60 km (5 km) thick ice shell. In contrast, the *N-S* basins require anti-roots of ~3 km or density anomalies of 6-80 kg/m$^3$ to explain their average 0.4 km depths. In the following paragraphs, we present a possible scenario to explain the distribution of Enceladus' basins: (i) the circumglobal *E*-basins originate from Pratt isostasy associated with viscous compaction above upwelling convective plumes beneath a thick stagnant lid along the paleo-equator and (ii) the antipodal *N-S* basins are formed by a combination of Airy and Pratt isostatic adjustments associated with ice shell thinning, cold convective downwellings associated with sluggish lid convection, and/or liquid water within the ice shell at the paleo-poles.

Convective upwellings beneath a thick stagnant lid are an effective process to form basins, where locally enhanced heating will reduce the near-surface porosity through viscous annealing and cause a positive density anomaly (Besserer et al. 2013). Thermal convection in Enceladus' ice shell is expected to be modulated primarily by the surface and ice-ocean interface temperatures, the yield strength of the lithosphere, the amplitude and distribution of tidal heating within the ice shell, as well as its thickness and rheology. At low latitudes where surface temperatures peak and tidal effects (heating and mechanical weakening of brittle ice) are relatively weak, convection with a thick, stagnant upper lid is predicted to occur if the ocean has nearly frozen out (Fig. 4). Moreover, linear stability analysis and numerical simulations of stagnant lid convection show that both the number (Kameyama et al. (2013) obtain a critical wavenumber of 12 in a 0.8 aspect ratio spherical shell) and the cylindrical shape of upwellings that may be expected (e.g., Solomatov 1995) are roughly consistent with the number and shape of the *E* basins. In further support of this hypothesis, a stagnant lid protects the surface from direct convective modification such that the cratered terrains along the paleo-equator should remain intact as observed. Moreover, relaxation of these craters implies high paleo heat fluxes (Bland et al. 2012), potentially consistent with this convective interpretation as well. We further note that mapping of Enceladus' fractures and ridges suggests that the satellite's shell experienced episodes of freezing and melting of its history, and that it was thicker in the past and thinned only recently (~200 Myr ago, Patthoff et al. 2016); thus, present-day (mean) ice shell thickness estimates of ~20 km (Thomas et al. 2016; Čadek et al. 2016; Van Hoolst et al. 2016; Beuthe et al. 2016) do not necessarily impact this hypothesized basin origin.

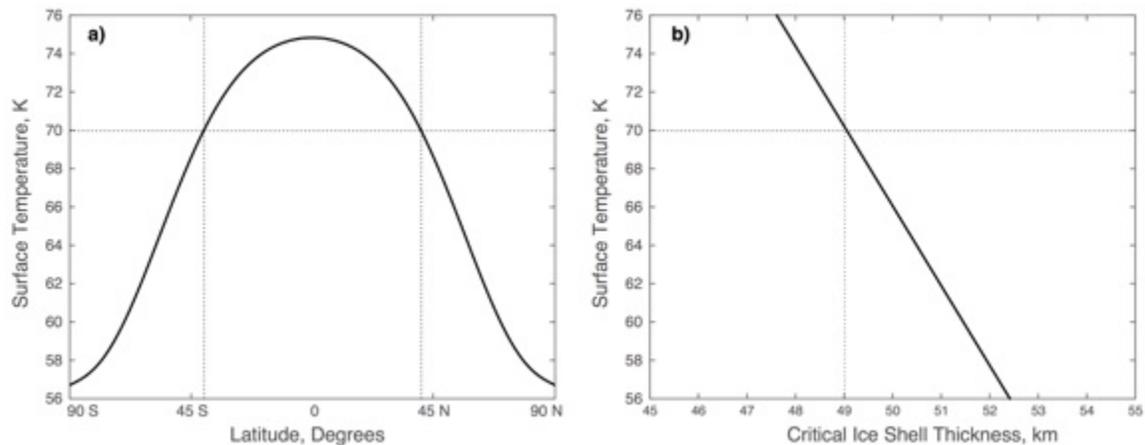

**Figure 4.** (a) Surface temperature as a function of latitude following Ojakangas & Stevenson (1989). (b) Critical ice shell thickness for convection to occur as a function of surface temperature assuming a nominal grain size of 0.1 mm (i.e., viscosity of $6 \cdot 10^{12}$ Pa s), melting temperature of 273 K, and Newtonian rheology following Barr & McKinnon (2007). If the ice shell is 49 km thick, convection is expected to occur where the surface

temperature is warmer than 70 K, or between 40°S and 40°N; here, a density anomaly of $\Delta\rho=19$ kg/m$^3$ is required to form the deepest (1.0 km) *E*-basin (Fig. 5b). This value requires a moderate 7% reduction in porous layer thickness for an initial porosity of 0.3.

Figure 4 implies that the onset of convection requires a thicker ice shell (or, conversely, smaller grain sizes) near the poles due to their colder surface temperatures. However, these calculations do not take tidal heating or brittle failure due to tidal stresses into account. Tidal heating is found to relax the grain size requirements for the onset of convection (e.g., Běhounková et al., 2013), while the production of any meltwater would reduce the ice viscosity by attenuating the internal stress field in the ice crystals (e.g., De La Chapelle et al. 1999; Kalousova et al. 2016). Conversely, tidal stresses are anticipated to reduce the yield strength of the lithosphere (e.g., Hammond & Barr 2015; Hammond et al. 2017) and promote sluggish lid convection or episodic, catastrophic overturning events (achieved by including plasticity and/or reducing the effective viscosity contrast across the shell in many numerical models; e.g., Han et al. 2012; Showman & Han 2005; Showman et al. 2013; Rozel et al. 2014; Barr & Hammond 2015). Ice shells as thin as 5 km can develop this type of vigorous convection beneath a thin, deformable lid if the viscosity contrast across the shell is less than three orders of magnitude (Barr, 2008). In this regime, convection is characterized by linear bands of upwelling plumes and large, cylindrically shaped downflows (e.g., Ratcliff et al. 1997). Since this cold, sinking ice is denser than its surroundings, we hypothesize that sluggish lid convection near the poles contributed to the formation of the *N-S* basins. Given the maximum warm-cold ice density contrast of $\Delta\rho=935-917=18$ kg/m$^3$, this mechanism would be viable for ice shells thicker than ~20 km (Fig. 3b) and if the thermal contrast remains to present day. For an ice shell $D > 20$ km thick, the thermal diffusion time is $\tau = D^2/\kappa > 10$ Myr assuming a thermal diffusivity of $\kappa = 1.26 \cdot 10^{-6}$ m$^2$/s, which is roughly consistent with the c. 10 Ma to 2 Ga estimates of the mid-latitudes on the leading and trailing hemispheres by Kirchoff & Schenk (2009). This requirement may be relaxed if compositional density contrasts are also present, which could arise if the regions of warm ice are relatively clean of low-eutectic impurities due to drainage of brines through the ice shell (Pappalardo & Barr 2004).

In contrast, basins are not expected to form at paleo-mid-latitudes due to the combination of colder surface temperatures and reduced tidal effects. As the surface temperature decreases away from the equator, the thickness of the stagnant lid may increase and limit the efficiency of near-surface viscous annealing by any upwelling plumes. Conversely, tidal strain rates decrease away from the poles such that surface fatigue, and any associated geologic activity, may not occur. The lack of significant topography between the *E* and *N-S* basins supports this hypothesis.

A similar pattern of equatorial and polar basins may then be expected to occur in Enceladus' current orientation. Indeed, the South Pole has a large topographic depression coincident with high heat flow and significant geologic activity (Porco et al., 2006, Spencer et al. 2006); the North Pole exhibits a very different morphology, however, and the origin of this dichotomy remains unclear. However, no additional basins are anticipated at mid- and low-latitudes since estimates of present-day ice shell thickness are too thin to drive convection (assuming linear stability analysis). This inferred reduction in ice shell thickness at present may be due to an episodic enhancement in tidal heating (Wisdom, 2004) and is consistent with thermal evolution models (Mitri & Showman, 2005).

**6. Surface geomorphology**

A stagnant lid at the paleo-equator would have protected the surface from convective modification, while a thin sluggish lid at the paleo-poles could have caused terrain modification. Both convective regimes show, in general, good agreement with the overall arrangement of geological provinces discussed above; for example, the resurfaced leading and trailing hemispheres, which used to be at the paleo-poles (Fig.2b), are superficially similar to the SPT, but differ significantly in the overall placement and nature of their terrain sub-units (Crow-Willard & Pappalardo, 2015). However, among the geological features in the provinces are structures located in the geologically named "Transitional" terrain (Crow-Willard & Pappalardo, 2015) (at $S1$-$S2$, and $\bar{S}$) such as peculiar ropy folds, called *funiscular plains*, materials that are otherwise found exclusively in the SPT region (Crow-Willard & Pappalardo, 2015; Spencer et al. 2009; Helfenstein et al. 2010; Nahm & Kattenhorn, 2015), hinting at past possible activity in Enceladus' paleo-poles. These are discussed in greater detail below. On the other hand, Enceladus' cratered terrains that were gathered mostly around the paleo-equator (Fig. 2a) hint at a past inactive lid.

It is reasonable to expect that volcanic and tectonic processes analogous to those active in the SPT may have accompanied the formation of paleo-polar depressions. However, the extent to which possible ancient analogs to these features are observed is disputed (Crow-Willard & Pappalardo, 2015; Pathoff 2015). Tectonic overprinting is widespread on Enceladus and recognizable early examples may largely have been destroyed. On the other hand, the traces of relict fracture patterns may be preserved as a result of cyclic reactivation from persistent tidal flexing, but it is unclear how the three-dimensional expression of the ancient features would degrade over time and under progressively evolving stress regimes.

Among potentially diagnostic morphological features that are associated with the active SPT are funiscular or ropy plains terrain.and narrow ridges called Dorsa are found throughout the SPT region, sometimes in polygonal networks.

Beyond the boundary of the modern SPT province, examples of morphological features that might be diagnostic of inactive cryovolcanic and tectonic processes associated with SPT style activity are rare and often not uniquely interpretable. They are absent along the paleo-equatorial band (Fig. 2, E1-E8) where the present morphological features are contiguous with those on adjacent terrains. Despite their rarity, a number of tentative examples have been identified. Fig. 5 shows four types of features that can be found on the trailing hemisphere of Enceladus (Helfenstein et al. 2010). These include putative examples of funiscular (ropy) ridges and shallow troughs the formation of which might have been shaped by cryovolcanic eruptions.

SPT analog features near or within south paleo-pole latitudes are generally more difficult to identify. A prominent polygonal arrangement of narrow ridges and domes (Ebony Dorsum and Cufa Dorsa) was known prior to the discovery of similar features near the south pole (Fig. 5, C1,2). A putative analog to an extinct tiger stripe segment (Fig. 4, D1,2) lies nearby the dorsa complex (Fig. 2, T2). Fig. 5 (B1,2) shows comparable bounding scarps that enclose examples of complexly-fractured reticulated plains. The bounding scarp B1 defines an edge to the SPT province. Fig. 2 (labeled B1) shows that in paleo-coordinates, the analog feature crosses into paleo-south-polar latitudes and its adjacent reticulated plains lies closer to the pole in the same sense as the SPT reticulated terrain lies poleward of the South Polar bounding scarp. Poleward of the paleo-reticulated plains are a set of long parallel fractures spaced about 10 km apart that branch in the Saturn-facing hemisphere in a pattern that is similar to the way that the tiger stripes Damascus and Baghdad branch in the modern SPT. These fractures lack the raised flanks of tiger stripes and they are spaced much more closely than the typical 35 km spacing of tiger stripes. However, the pattern of stresses that created the branch pattern preferentially on the Saturn-facing hemisphere may be similar to that which created the branching of the tiger stripes.

Geomorphological evidence offers few clues about the time scales over which the polar shift occurred. While a number of localized ancient features appear to exist in the modern day leading- and trailing-hemispheres, it is clear that the large-scale arrangement of tectonic features preclude their strictly being ancient clones of the current SPT configuration (Crow-Willard and Pappalardo 2015). The existence of localized examples of features in the Leading- and Trailing Hemispheres that might have been created by similar physical processes to features in the SPT region, for example, troughs of any length and orientation that are bounded by shallow ridges (possibly created by fissure eruptions in a manner similar to the way tiger stripe flanks are formed around the erupting medial fracture), or vermicular arrangements of rounded ridges that resemble SPT funiscular plains materials simply suggest that conditions on what is now the leading and trailing hemispheres were, at one time right to cause surface expressions of volcanism and folding, albeit in a different geological arrangement than the SPT. The time scale over which these conditions prevailed is unknown, but the fact that these hemispheres appear to have never fully evolved the same arrangement of tectonic features as the SPT suggests that there may have been episodic changes in the extent and intensity of subsurface hot spots since the onset of the migration. Kirchoff and Schenk (2009) concluded that recent polar wander had not occurred because the present-day crater density pattern is symmetrical about the present-day equator. Because the rotation likely happened about an axis that is nearly coincident with the tidal axis, the present leading-trailing asymmetry may have originated as a northern-southern hemisphere asymmetry, which would be preserved only if the polar wander event occurred on a relatively short time scale.

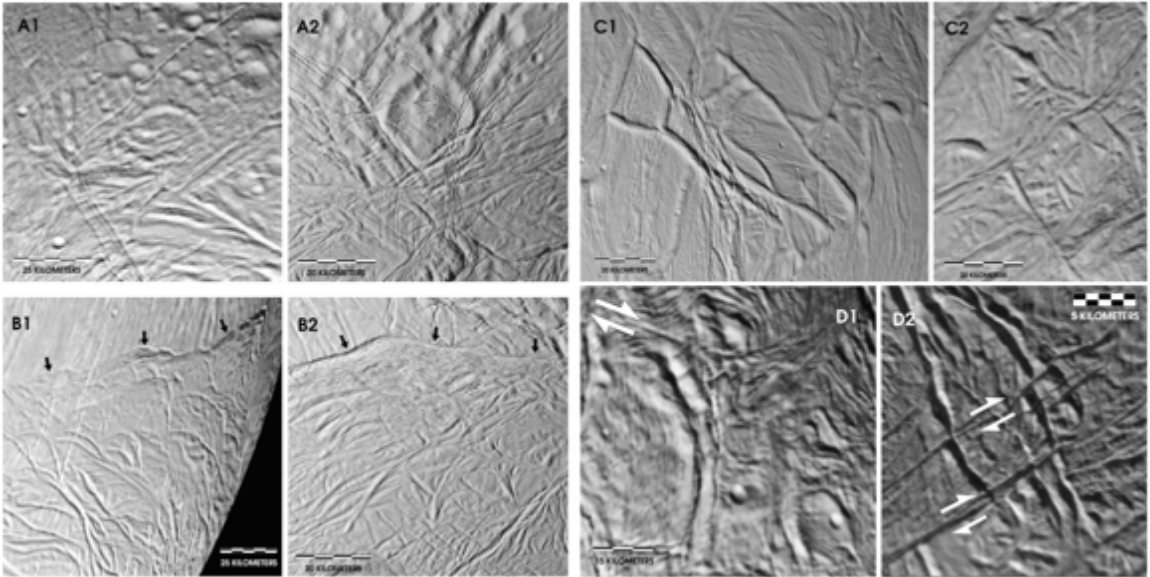

**Figure 5.** Comparison of proposed relict SPT-style features in Enceladus' Eastern Hemisphere with SPT examples. A1-A2 shows thermally eroded, laterally deformed craters that suggest regions bounding Sarandib Planitia (A1 at 177°W, 50°S) and the SPT (A2 at 316°W, 38°N) and in the Eastern hemisphere have undergone very similar thermal histories and margin deformation. B1-B2 compares (B1) south facing scarp and enclosed reticulated plains at about 15°S in the Eastern hemisphere with (B2) a similar bounding scarp and enclosed reticulated plains in the SPT (200°W, 55°S). C1-C2 compares the well-known Eastern Hemisphere dorsa network (C1) containing Ebony Dorsum (281°W, 6°N) and Cufa Dorsa (286°W, 3°N) with (C2) a similar unnamed dorsa network within the South Polar reticulated plains (320°W, 73°S). D1-D2 compares (D2) morphologically degraded relict tiger stripe features at the end of Damascus Sulcus (0°W, 76°S) with (D1) morphologically similar example in the Eastern hemisphere within Diyar Planitia (277°W, 20°N). Arrows in D1-D2 identify right-lateral strike-slip offsets. Adapted from Spencer et al. (2009).

## 7. Conclusion

We built the shape of Enceladus represented by a series of spherical harmonic coefficients up to degree 16 using updated limb profiles and control points obtained from Cassini ISS images. We discovered sets of basins that follow a global pattern: a pair of four antipodal basins orthogonal to eight basins that follow a great circle. We argue that these basins are remnants of paleo-poles and a paleo-equator and that Enceladus experienced a True Polar Wander event in the past that placed these basins at their current locations. At the paleo-equator where convection took place beneath an icy thick stagnant lid that led to local reduction in porosity, local positive density anomalies may have been compensated through Pratt isostasy and could have formed the *E* basins. On the other hand, cold, dense convective downwellings surrounded by warm upwelling ice beneath a thin sluggish lid may have led to density contrasts that could have formed the *N-S* basins through Pratt isostasy. Although our interpretation is not unique, it shows consistency with Enceladus' surface geomorphology, where *E* basins generally follow the cratered line, while the S and N basins are located in the currently resurfaced trailing and leading hemisphere terrains.

From the spatial distribution of the observed basins and their possible geophysical interpretations, we argue that Enceladus likely experienced TPW. Such a phenomenon is expected to create a stress field and leave a tectonic record on the satellite's surface (Matsuyama et al. 2014; Matsuyama & Nimmo, 2008; Schenk et al. 2008). The formation of today's SPT and the resulting mass anomaly may have been the primary driver of the satellite's reorientation. We estimate the load required to cause such a reorientation by using Eq. (42) from Matsuyama & Nimmo (2007) (assuming Enceladus reorients about the tidal axis for simplification) with a final position of the mass anomaly being 8° from the current spin pole. We obtain a load Q in the range of -3.6 to -3, which is between the upper limit -2 suggested by Nimmo & Pappalardo (2006) and that suggested by Matsuyama & Nimmo (2007) of -4. An endogenic process such as diapirism is one way to form such a mass anomaly (Nimmo & Pappalardo, 2006); however, compositional convection rather than a thermal one (Stegman et al. 2009) is more likely to have occurred to account for such a large load; but it is still unclear how a diapir could form in that particular region (e.g, Rozel et al. 2014). In addition, Isostatic processes (like the ones we propose here to explain *E*, *N*, and *S* basins) are not the most ideal way to drive reorientation; first the compensation reduces the gravity anomaly, and second, those processes form gravity anomalies that are too small to cause such a large reorientation. Therefore, the process that formed the SPT before the reorientation is probably a different than those that formed the *E*, *N*, and *S* basins. Early placement of the SPT on the leading hemisphere is significant because preferential heavy bombardment there (Zahnle et al. 2001), especially by a large basin-forming impactor, could create a potential anomaly (Nimmo & Matsuyama, 2007) and is a possible trigger for warm diapirism (Ghods & Arkani-Hamed, 2007; Roberts & Arkani-Hamed, 2009). Thus, we speculate that the activity in the SPT may have been initiated by an impact. In any case, it is peculiar that Enceladus is the only satellite in the Saturnian system that does not have a large impact basin.

The polar asymmetry seen today remains peculiar; nonetheless, our results show that Enceladus before TPW had polar symmetry, with topographic and geological features that can be explained through plausible geophysical processes. The combination of topography, surface geomorphology, and the current polar dichotomy on Enceladus, all together, make the TPW hypothesis very plausible.

# Appendix A

## Uncertainties.

This work presents two types of results: the first is the topographic map, and the second is its interpretation as evidence of TPW; each has uncertainties.

## A1. Uncertainties in topography

In order to estimate the uncertainty on Enceladus' topography, we apply two different methods. The first is straightforward and consists of applying the weighted least squares method to fit the coefficients of the spherical harmonics function to the observations. Thus each data point (limb or control point) was weighted by its estimated observation uncertainty. The square roots of the diagonal elements of the inverted covariance matrix represent the uncertainties on the estimated coefficients.

In the second method, we used the Bootstrapping technique, where we created synthetic data, but added Gaussian noise to each (X, Y, Z) measurement of a data point with a sigma equal to the estimated uncertainty of that point. After disturbing all of the points by their own uncertainties, a new fit of spherical harmonic coefficients was done. The process was repeated over 1000 iterations, which produced a Gaussian distribution of the possible solutions for each coefficient; the mean values and standard deviations of those distributions represent the coefficients' final fitted values and uncertainties.

Once estimated, the uncertainties on the coefficients are converted to topography by propagation of uncertainties as follows,

$$\sigma_r(\theta,\phi) = R_0 \sqrt{\sum_{l=0}^{N}\sum_{m=0}^{l} P_{lm}^2(\cos\theta)\left[\sigma_{C_{lm}}^2 \cos(m\phi)^2 + \sigma_{S_{lm}}^2 \sin(m\phi)^2\right]}, \tag{A1}$$

where $\sigma_{C_{lm}}$ and $\sigma_{S_{lm}}$ are uncertainties on $C_{lm}$ and $S_{lm}$, respectively. Both methods result in the same uncertainty map presented in $3\sigma$ in Fig. A1 showing that uncertainties in topography varies between ~75 and ~100 m. Hence, all the basins studied in this paper are statistically significant. Here, we did not take into account any systematic errors. While they are present, their effects are relatively small compared to the random ones. Towards this end, we plot in Fig. A2 a histogram of the residuals of the data points to the fitted shape. The Gaussian distribution of the post-fit errors means that the sources of error have more of a random behavior rather than a systematic.

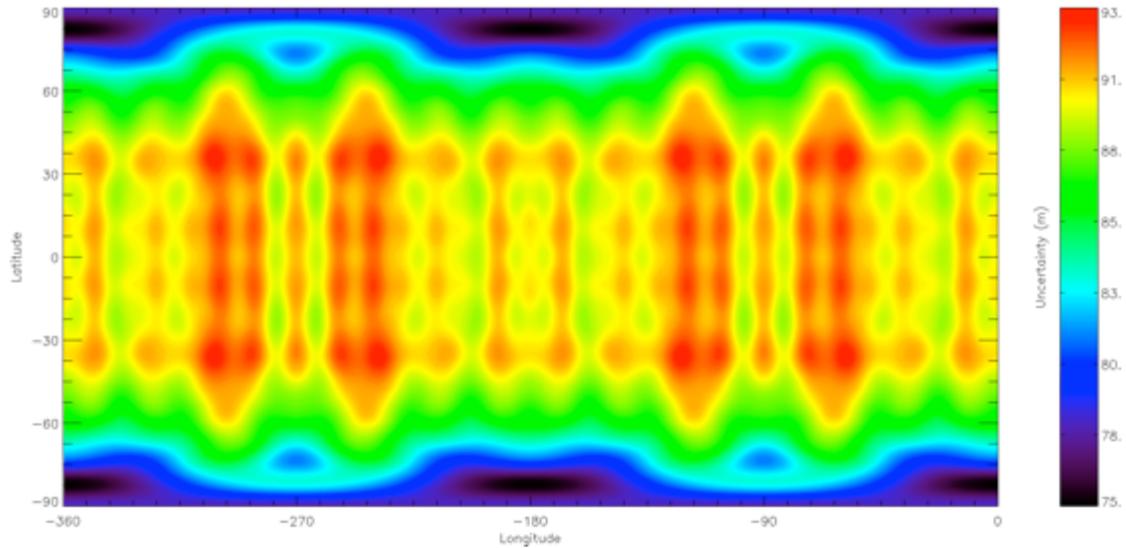

**Figure A1.** Uncertainty map on Enceladus' topography showing 3σ uncertainties ranging from 75 to 93 m.

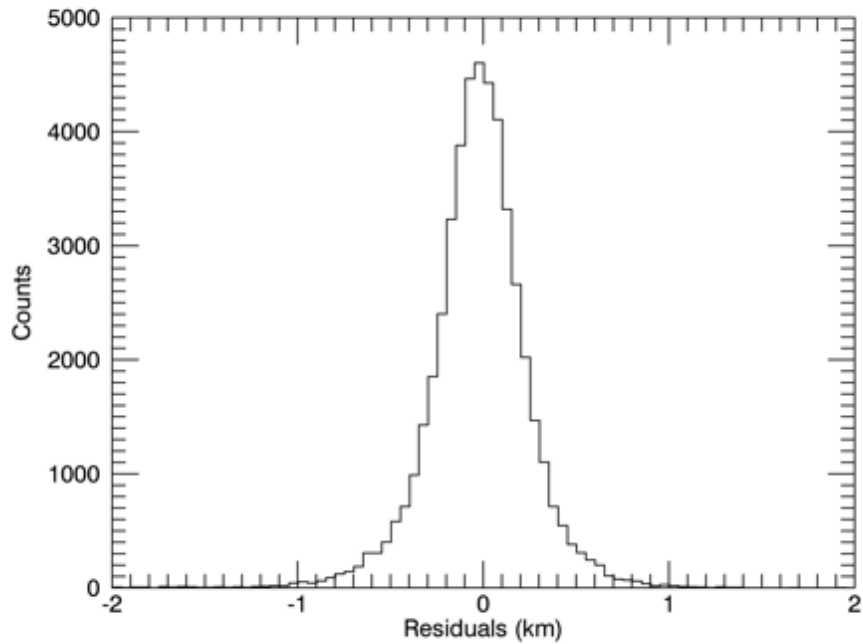

**Figure A2.** Histogram of Post-fit residuals showing a Gaussian distribution of error.

**A2. Uncertainties in TPW estimations.**

To estimate the TPW reorientation angle and axis, and their uncertainties, we used the coordinates of $E$ basins given in Table 1 and their widths as uncertainties to perform a weighted least square method to fit the plane. As previously, the uncertainties on the plane characteristics have been obtained from the diagonal terms of the covariance matrix and converted to uncertainties in paleo-polar positions (Fig. 1b). Note that our TPW characterization is based on the observed basin locations and widths, without taking into consideration any geophysical phenomena that might have affected them. For instance, the formation of the topographic low at Enceladus' SPT is very likely to have led to the formation of the elevated terrain surrounding that area around latitude 50°S (Schenk & McKinnon, 2009) (Figs. C1, C2). Since the SPT is the youngest terrain on Enceladus, it may have consequently erased some of the older basins (if they existed) near that region, or moved their center (i.e. basins E7, and S1-3).

**Appendix B**

**Coefficients of the spherical harmonics function**

We attached two text files to this article (C_lm.txt and S_lm.txt) listing the coefficients $C_{lm}$ and $S_{lm}$ of the fitted spherical harmonics function (Eq. 1). The coefficients used in this work are not normalized. To see the contribution in kilometres of each coefficient to satellite's shape one has to simply multiply it by the mean radius $R_0 = 252.22$ km.

The associated Legendre polynomials are define as (for $m \geq 0$)

$$P_l^m(x) = (-1)^m \left(1 - x^2\right)^{m/2} \frac{d^m}{dx^m} P_l(x). \tag{B1}$$

The columns and lines represent the $l$ and $m$ numbers. Many of those coefficients are set to zero; those are inexistent coefficients (e.g. $m > l$).

**Appendix C**

**Data coverage and additional maps**

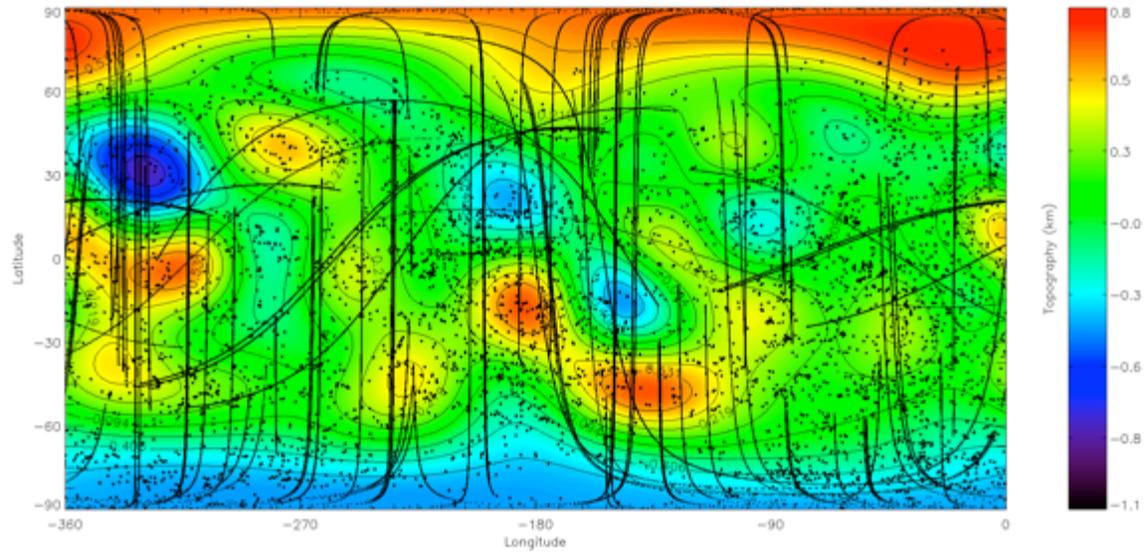

**Figure C1.** Topographic map of Enceladus based on the l=3 to 8 terms in a spherical harmonic expansion of the satellite's shape. The map is similar to the one made by Nimmo et al. 2011. Lines crossing the map represent data from limb profiles, and dots represent surface control points.

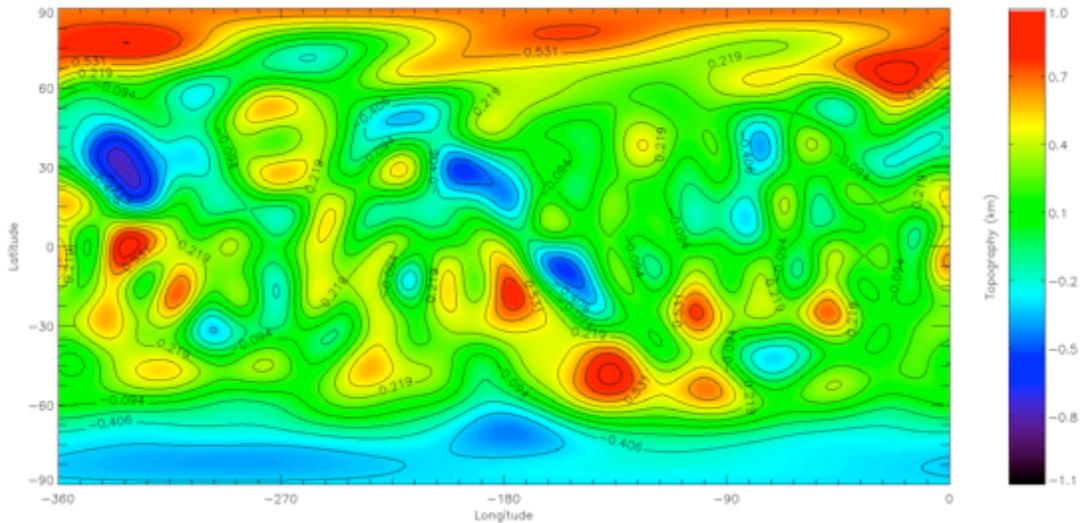

**Figure C2.** Topographic map of Enceladus based on the l=3 to 16 terms in a spherical harmonic expansion of the satellite's shape. The l=3 feature representing the polar asymmetry dominates the satellite's global topography. Although the E, S, and N features are visible on the map, the removal of the terms representing the polar asymmetry is necessary for a better visualization of those features.

**Appendix D**

**Uniqueness of basin arrangement**

The proposed E basin arrangement on a great circle and antipodal N/S locations was detected by eye; thus, one might question whether such an arrangement is purely random, and that a similar arrangement could be drawn from another set of basins. We test the significance of our detection by choosing different positions of poles. Therefore, we tested pole longitudes between 0° and 360° and colatitudes between 0° and 90° (limited to this range in colatitude because of polar symmetry in the tested great circles), with an iteration step of an arc-distance of ~22 km ($\Delta\theta=5°$) between the tested poles (861 iterations in total). To assure an equidistant distribution of the tested poles we divide the satellite's surface into circles with constant colatitudes separated by $\Delta\theta=5°$. In this case each circle can have $N_i$ longitudes of poles, such as

$$N_i = \frac{2\pi}{\Delta\theta}\sin\theta_i \tag{D1}$$

where $\theta_i$ is the colatitude of the *i*th circles. Thus, the angular separation between longitudes is $\Delta\phi_i = 2\pi/N_i$, and the coordinates of the *j*th point on the *i*th circle is ($i\Delta\theta$, $j\Delta\phi_i$). The aim of this test is to verify whether there are other configurations that could: 1) Produce an alignment of basins with comparable or lower mean topography along the tested great circle, and 2) place other basins at both poles of that great circle.

For each pole position, we calculate the mean topography along the tested great circle, and then search for mean topography extrema within ±10° of latitude (approximately the size of a basin). Some rotations did not return any extrema within the imposed limits, others returned topographic minima, maxima, or both. The distribution of the results of the simulation is shown in Figure D1; most of the mean topography extrema are gathered around an average zero, with a peak near ~30-40 m and one-sigma scatter of ~60 m. We investigate cases where basin alignments have mean topography minima along the great circle below -120 m (2-sigma). The remaining solutions can be divided into three sets of basins that align on three different great circles (many great circles can pass through the same set because of the basin depth). From each set, we select the solution that results in the lowest mean topography along the great circle and plot in Figure D2a, the trajectory of the great circle on top of the topographic map (note that the basin at 230°W, -15°N, Fig. 1, is not part of any of these three sets). In the left panels of Figure 4b-d, we place the great circle 'b' at the equator by rotating the map -20° about the Z-axis then -75° about the X-axis (Fig. D2b); the great circle 'c' at the equator through similar rotations of 10° then 55° (Fig. D2c); and the great circle 'd' at the equator by rotating the map by 70° then -40° (Fig. D2d).

The right panels of Figure 4b-d show plots of the mean topography versus latitude in the reoriented coordinates. Map 'c' (Fig. D2c) is the only one that has mean topography minima at both the equator and poles. Moreover, the maps 'b' and 'd' (Figs. D2b, D2d) have a minimum mean topography in the equatorial region of about -140 m and -150 m, respectively, which are inside the three-sigma scatter of topography extrema distribution (Fig. D1); while the lowest mean topography along the equator of the map 'c' (Fig. D2c) is about -240 m, ~100 m deeper on average than the that of the other two cases, and is the only case with topography extrema outside of the 3-sigma scatter (Fig. D1). We, therefore, base the TPW hypothesis on map 'c' and conclude that our eye-detected great circle passing approximately through the *E*-basins is unique.

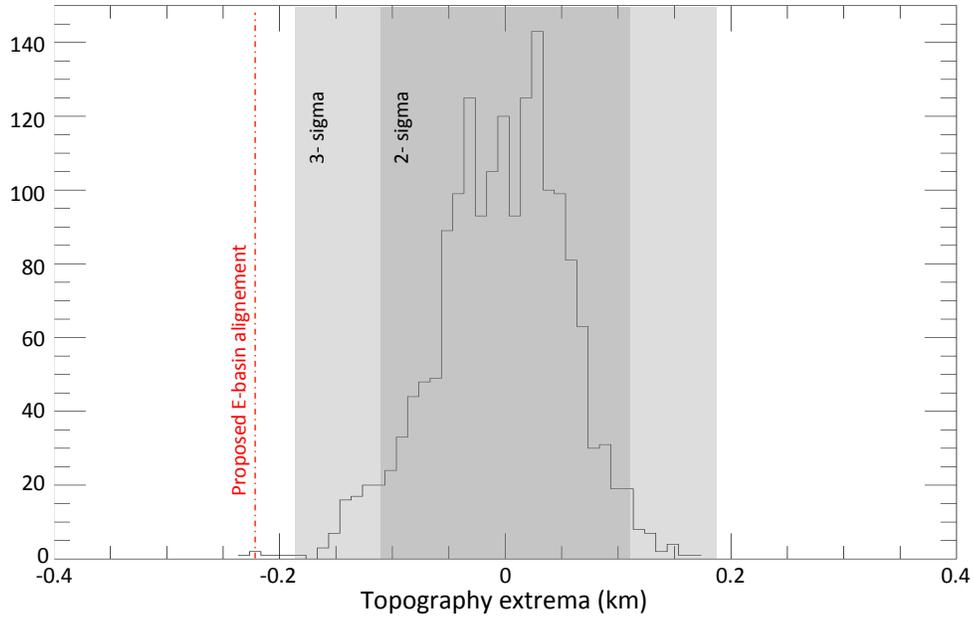

**Figure D1.** Histogram of topography extrema along all of the possible great circles on Enceladus (with an arc-distance of 5° between each great circles). The great circles passing through the *E* basins are the only ones that have a mean elevation beyond the 3-sigma distribution.

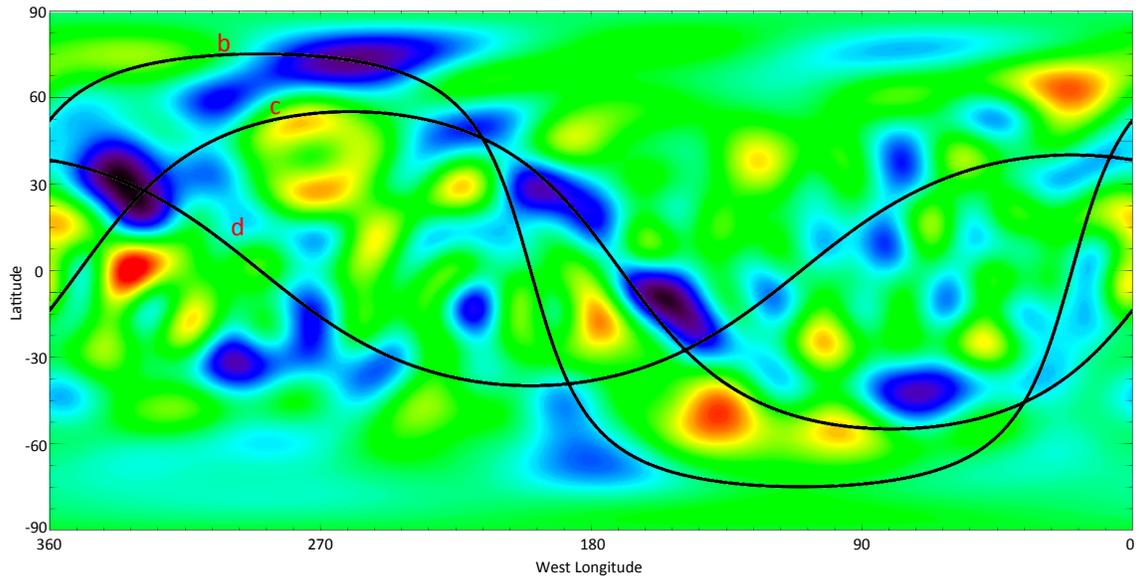

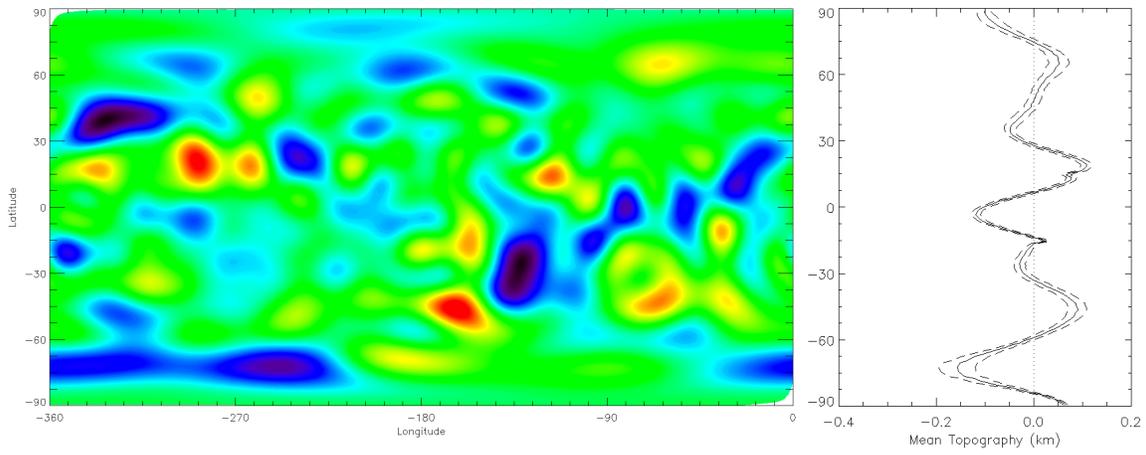

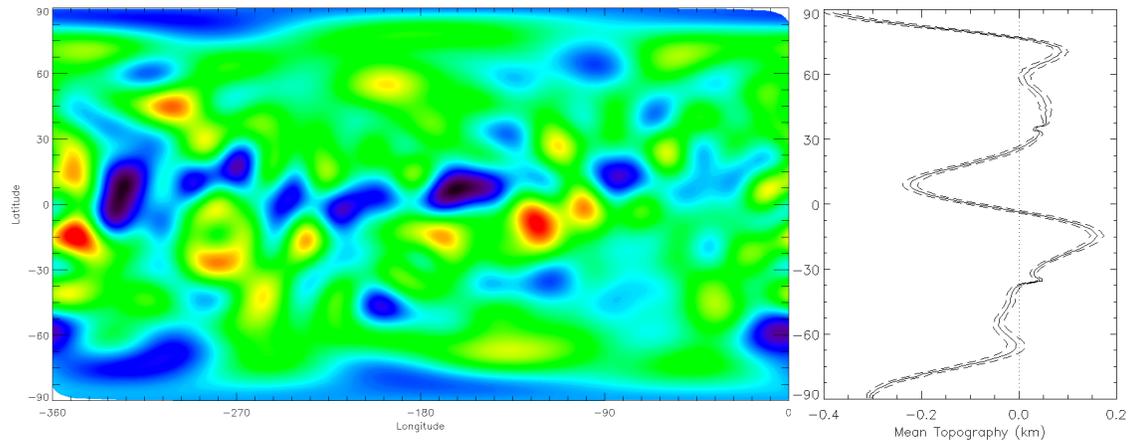

d

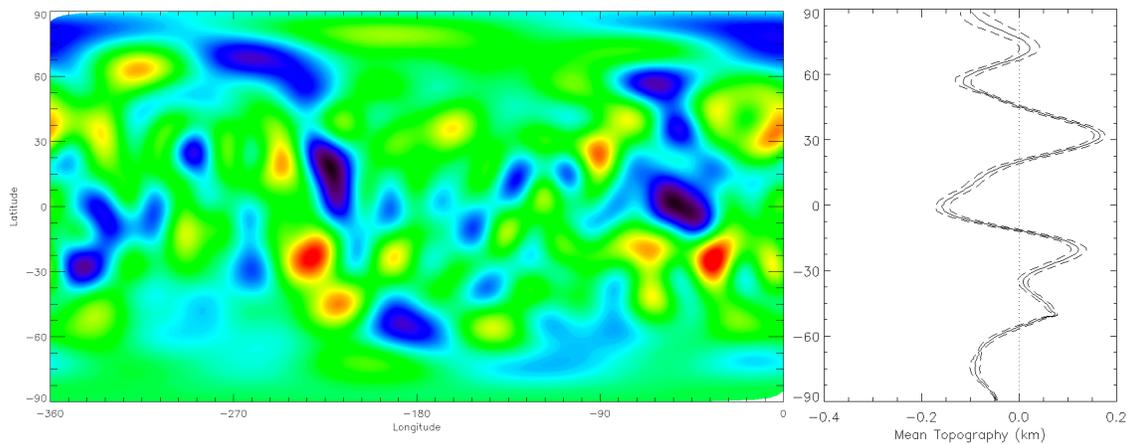

**Figure D2.** Solutions for great circles that pass through different sets of basins. (a) Great circles that have a mean depth below the 2-sigma level (Fig. 3). (b-d) Left panels: reorientation of Enceladus' map to place those great circles at the equator; right panels: mean topography versus latitude for each reoriented map. Map (c), the arrangement that our TPW hyupothesis is based on, has the lowest equatorial mean topography and is the only one that has topographic lows at its poles.

**References:**
Barr, A.C. & Mckinnon, W.B., Convection in Enceladus' ice shell: Conditions for initiation. *Geophys. Res. let.*, **34**, 9, CiteID L09202 (2007).
Barr, A.C. Mobile lid convection beneath Enceladus' south polar terrain, *J. Geophys. Res. Planets*, **113**, Issue E7, CiteID E07009 (2008).
Barr, A.C. & Hammod, N.P. A common origin for ridge-and-trough terrain on icy satellites by sluggish lid convection, *Physics of the Earth and Planetary Interiors*, **249**, 18-27 (2015).
Besserer, J., Nimmo, F., Roberts, J.H. & Pappalardo, R.T. Convection-driven compaction as a possible origin of Enceladus's long wavelength topography, *J. Geophys. Res. Planets*, **118**, Issue 5, 908-915 (2013).
Bouley, S., Baratoux, D., Matsuyama, I. et al. Late Tharsis formation and implications for early Mars. *Nature*, **531**, Issue 7594, 344-347 (2016).
Čadek, O., Tobie, G., Van Hoolst, T. et al. Enceladus's internal ocean and ice shell constrained from Cassini gravity, shape, and libration data. *Geophysical Research Letters,* **43**, Issue 11, 5653-5660 (2016).
Cooper, N.J., Murray, C.D., Lainey, V. et al. Cassini ISS mutual event astrometry of the mid-sized Saturnian satellites 2005-2012, *A&A*, **572**, id.A43, 8 (2014).
Crow-Willard, E.N. & Pappalardo, R.T. Structural mapping of Enceladus and implications for formation of tectonized regions, *J. Geophys. Res.*, **120**, Issue 5, 928-950 (2015).
Davies, M.E., Colvin, T.R., Oberst, J. et al. The control networks of the Galilean satellites and Implications for global shapes, *Icarus*, **135**, Issue 1, 372-376 (1998).

De La Chapelle, S., Milsch, H., Castelnau, O., & Duval, P. Compressive creep of ice containing a liquid intergranular phase: Rate-controlling processes in the dislocation creep regime, *Geophys. Res. Lett.*, **26**, Issue 2, 251–254 (1999).

Dermott, S.F. & Thomas, P.C. The shape and internal structure of Mimas, *Icarus*, **73**, 25-65 (1988).

Deschamps, F., Tackley, P.J., & Nakagawa, T. Temperature and heat flux scalings for isoviscous thermal convection in spherical geometry, *Geophysical Journal*, **182**, Issue 1, 137-154 (2010).

Doubrovine, P. V., Steinberger, B. & Torsvik T. H. Absolute plate motions in a reference frame defined by moving hot spots in the Pacific, Atlantic, and Indian oceans, *J. Geophys. Res.*, **117**, B09101, doi:10.1029/2011JB009072 (2012).

Edwards, K. Geometric Processing of Digital Images of the Planets, *Photogrammetric Engineering and Remote Sensing*, **53**, 1219-1222 (1987).

Giese, B. & Cassini Imaging Team. The Topography of Enceladus, *EPSC*, p.675 (2010).

Ghods, J. & Arkani-Hamed, J. Impact-induced convection as the main mechanism for formation of lunar mare basalts, *J. Geophys. Res.*, **112**, Issue E3, citeID E03005 (2007).

Glein, C.R. et al. The pH of Enceladus' ocean, *Geochimica et Cosmochimica Acta*, **162**, 202-219 (2015).

Hammond, N.P., Barr, A.C., Hirth, G., & Cooper, R.F. The weakening or strengthening of water ice in response to cyclic loading, In *Lunar and Planetary Science Conference,* **Vol. 48** (2017).

Han, L., Tobie, G., & Showman, A. P. The impact of a weak south pole on thermal convection in Enceladus' ice shell. *Icarus*, **218**, Issue 1, 320-330 (2012).

Helfenstein et al. Leading-Side Terrains on Enceladus: Clues to Early Volcanism and Tectonism from Cassini ISS, *American Geophysical Union*, Fall Meeting, abstract #P23C-04 (2010).

Kalousová, K., Souček, O., Tobie, G., Choblet, G., & Čadek, O. Water generation and transport below Europa's strike-slip faults. *J. Geophys. Res. Planets,* **121**, Issue 12, 2444-2462 (2016).

Kirchoff, M. R., & Schenk, P. Crater modification and geologic activity in Enceladus' heavily cratered plains: Evidence from the impact crater distribution. *Icarus*, **202**, Issue 2, 656-668 (2009).

Matsuyama, I., Nimmo, F. & Mitrovica, J.X. Planetary Reorientation, *Annual Review of Earth and Planetary Sciences*, **42**, 605-634 (2014).

Matsyuama, I. & Nimmo, F. Tectonic patterns on reoriented and despun planetary bodies, *Icarus*, **195**, Issue 1, 459-473 (2008).

Mitri, G., & Showman, A. P. Convective–conductive transitions and sensitivity of a convecting ice shell to perturbations in heat flux and tidal-heating rate: Implications for Europa. *Icarus*, **177**, Issue 2, 447-460 (2005).

Mitrovica, J.X. & Wahr, J. Ice Age Earth Rotation, *Annual Review of Earth and Planetary Sciences*, **39**, 577-616 (2011).

Nahm, A.L. & Kattenhorn S.A. A unified nomenclature for tectonic structures on the surface of Enceladus, *Icarus*, **258**, 67-81 (2015).

Nimmo, F. & Pappalardo, R.T. Diapir-induced reorientation of Saturn's moon Enceladus, *Nature*, **441**, Issue 7093, 614-616 (2006).

Nimmo, F., Thomas, P.C., Pappalardo, R.T. & Moore, W.B. The global shape of Europa: Constraints on lateral shell thickness variations. *Icarus*, **191**, Issue 1, 183-192 (2007a)**.**
Nimmo, F. & Matsuyama, I. Reorientation of icy satellites by impact basins, *Geophys. Res. Let.*, **34**, Issue 19, CiteID L19203 (2007b).
Nimmo, F., Bills, B.G., Thomas, P.C. & Asmar, S.W. Geophysical implications of the long-wavelength topography of Rhea, *J. Geophys. Res.*, **115**, Issue E10, CiteID E10008 (2010).
Nimmo, F., Bills, B.G., Thomas, P.C. Geophysical implications of the long-wavelength topography of the Saturnian satellites, *J. Geophys. Res.,* **116**, Issue E11, CiteID E11001 (2011).
Oberst, J., Zubarev, A., Nadezhdina, I., Shishkina, L., Rambaux, N. The Phobos geodetic control point network and rotation model, *Planet. & Space Sci.*, **102**, 45-50 (2014).
Ojakangas, G.W. & Stevenson, D.J., Thermal state of an ice shell on Europa, *Icarus*, **81**, 220-241 (1989).
Owen, W. M. Jr.  Cassini ISS geometric calibration, *JPL Interoffice Memorandum*, **312**, E-2003-001 (2003).
Patthoff, D.A. and Kattenhorn S.A. A fracture history on Enceladus provides evidence for a global ocean. *Geophys. Res. Let.*, **38**, Issue 18, Cite L18201  (2011).
Patthoff D. A. et al. Exploring Enceladus' geologic history through time. *LPSC*, sess **705**, 1772 (2016).
Porco, C.C. et al. Cassini Observes the Active South Pole of Enceladus, *Science*, **311**, Issue 5766, 1393-1401 (2006).
Ratcliff, J.T., Tackley, P.J., Schubert, G. & Zebib, A. Transitions in thermal convection with strongly variable viscosity, *Physics of the Earth and Planetary Interiors*, **102**, Issue 3, 201-212 (1997).
Reese, C. C., Solomatov, V.S., Baumgardner, J.R. & Yang, W.S. Stagnant lid convection in a spherical shell, *Physics of the Earth and Planetary Interiors*, **116**, Issue 1-4, 1-7 (1999).
Roberts, J.H. & Arkani-Hamed, J. Impact-induced mantle dynamics on Mars, *Icarus*, **218**, Issue 1, 278-289 (2012).
Rozel, A., Besserer, J., Golabek, G.J., Kaplan, M. & Tackley, P.J. Self-consistent generation of single-plume state for Enceladus using non-Newtonian rheology, *J. Geophys. Res. Planets*, **119**, Issue 3, 416-439 (2014).
Siegler, M.A., Miller, R.S., Keane J.T. et al. Lunar true polar wander inferred from polar hydrogen. *Nature*, **531**, Issue 7595, 480-484 (2016).
Schenk, P.M., Matsuyama, I. & Nimmo, F. True polar wander on Europa from global-scale small-circle depressions, *Nature,* **453**, Issue 7193, 368-371 (2008).
Schenk, P.M. & McKinnon, W. B. One-hundred-km-scale basins on Enceladus: Evidence for an active ice shell, *Geophys. Res, Let.*, **36**, Issue 16, CiteID L16202 (2009).
Showman, A. P., & Han, L. Effects of plasticity on convection in an ice shell: Implications for Europa, *Icarus*, **177**, Issue 2, 425-437 (2005).
Showman, A. P., Han, L. & Hubbard, W. B. The effect of an asymmetric core on convection in Enceladus' ice shell: Implications for south polar tectonics and heat flux. *Geophys. Res. Lett.*, **40**, Issue 21, 5610-5614 (2013).

Solomatov, V.S., Scaling of temperature- and stress-dependent viscosity convection, *Phys. Fluids*, **7**, Issue 2, 266-274 (1995).

Spencer, J.R., Pearl, J.C., Segura, M. et al. Cassini Encounters Enceladus: Background and the Discovery of a South Polar Hot Spot, *Science*, **311**, Issue 5766, 1401-1405 (2006).

Spencer, J.R., Barr, A.C., Esposito, L.W. Enceladus: An Active Cryovolcanic Satellite, Saturn from Cassini-Huygens, *Springer Science Business Media B.V.*, 683 (2009).

Stegman, D.R., Freeman, J. & May, D.A. Origin of ice diapirism, true polar wander, subsurface ocean, and tiger stripes of Enceladus driven by compositional convection, *Icarus*, **202**, Issue 2, 669-680 (2009).

Tajeddine, R., Cooper N.J., Lainey, V., Charnoz, S., Murray, C.D. Astrometric reduction of Cassini ISS images of the Saturnian satellites Mimas and Enceladus, *A&A*, **551**, id.A129, 11 (2013).

Tajeddine, R., Rambaux, N., Lainey, V. et al. Constraints on Mimas' interior from Cassini ISS libration measurements, *Science*, **346**, Issue 6207, 322-324 (2014).

Tajeddine, R., Lainey, V., Cooper N.J., Murray, C.D. Cassini ISS astrometry of the Saturnian satellites: Tethys, Dione, Rhea, Iapetus, and Phoebe 2004-2012, *A&A*, **575**, id.A73, 6 (2015).

Thomas, P.C., Davies, M.E., Colvin, T.R. et al. The Shape of Io from Galileo Limb Measurements, *Icarus*, **135**, Issue 1, 175-180 (1998).

Thomas, P.C., Joseph, J., Carcich, B. et al. Eros: Shape, topography, and slope processes, *Icarus*, **155**, Issue 1, 18-37 (2002).

Thomas, P.C., Burns, J.A., Helfenstein, P. et al. Shapes of the Saturnian icy satellites and their significance, *Icarus*, **190**, 573-584 (2007).

Thomas, P.C. Sizes, shapes, and derived properties of the saturnian satellites after the Cassini nominal mission, *Icarus*, **208**, Issue 1, 395-401 (2010).

Thomas, P.C., Tajeddine, R., Tiscareno, M.S. et al. Enceladus's measured physical libration requires a global subsurface ocean, *Icarus*, **264**, 37-47 (2016).

Walker, C.C., Bassis, J.N., & Liemohn, M.W. On the application of simple rift basin models to the south polar region of Enceladus, *Journal of Geophysical Research: Planets*, **117**, E7, CiteID E07003 (2012)

West, R., Knowles, B., Birath, E. et al. In-flight calibration of the Cassini imaging science sub-system cameras, Planet. Space Sci., **58**, Issue 11, 1475-1488 (2010).

Wisdom, J. Spin-orbit secondary resonance dynamics of Enceladus, *The Astronomical Journal*, **128**, Issue 1, 484-491 (2004).

Yoshida, M. & Kageyama, A. Low-degree mantle convection with strongly temperature- and depth-dependent viscosity in a three-dimensional spherical shell, *J. Geophys. Res.*, **111**, Issue B3, CiteID B03412 (2006).

Zahnle, K., et al. Differential cratering of synchronously rotating satellites by ecliptic comets. *Icarus*, *153*, Issue 1, 111-129 (2001).

Zebker, H.A., Stiles, B., Hensley, S. et al. Size and Shape of Saturn's Moon Titan, *Science*, **324**, Issue 5929, 921 (2009).

**Acknowledgements.** The authors wish thank the anonymous reviewers for their suggestions that improved the quality of the paper. This work was mainly supported by the Cassini mission. KMS and PH gratefully acknowledge support from NASA OPR Grant NNX14AR28G and NASA CDAP Grant NNX12AG82G, respectively. Many thanks to Todd Ansty, Thorsten Becker, Lukas Fuchs, Alex Hayes, Jonathan Joseph, Phil Nicholson, Pam Smith, and Matt Weller for their assistance and fruitful discussions.